\documentclass[twocolumn,showpacs,amsmath,amssymb,aps,prd]{revtex4}

\usepackage{bm}

\begin{document}

\title{The quantum of area $\Delta A = 8\pi l_P^{2}$
and a statistical interpretation of black hole entropy}

\author{Kostiantyn  Ropotenko}
 \email{ro@stc.gov.ua}

\affiliation{State Administration of Communications, Ministry of
Transport and Communications of Ukraine,\\ 22, Khreschatyk, 01001,
Kyiv, Ukraine}

\date{\today}

\begin{abstract}
In contrast to alternative values, the quantum of area $\Delta A =
8\pi l_P^{2}$ does not follow from the usual statistical
interpretation of black hole entropy; on the contrary, a statistical
interpretation follows from it. This interpretation is based on the
two concepts: nonadditivity of black hole entropy and Landau
quantization. Using nonadditivity a microcanonical distribution for
a black hole is found and it is shown that the statistical weight of
black hole should be proportional to its area. By analogy with
conventional Landau quantization, it is shown that quantization of
black hole is nothing but the Landau quantization. The Landau levels
of black hole and their degeneracy are found. The degree of
degeneracy is equal to the number of ways to distribute a patch of
area $8\pi l_P^{2}$ over the horizon. Taking into account these
results, it is argued that the black hole entropy should be of the
form $S_{bh} =2\pi\cdot\Delta\Gamma $, where the number of
microstates is $\Delta\Gamma = A/8\pi l_P^{2}$. The nature of the
degrees of freedom responsible for black hole entropy is elucidated.
The applications of the new interpretation are presented. The effect
of noncommuting coordinates is discussed.

\end{abstract}


\maketitle

\section{Introduction}

The statistical source of the Bekenstein-Hawking black hole entropy
\begin{equation}
\label{ent0} S_{bh}=\frac{A}{4 l_P^{2}}
\end{equation}
is still a central problem in black hole physics. Quantization of
the black hole area can be one of the keys to understanding of it.
According to Bekenstein \cite{bek}, quantization of the black hole
area means that the area spectrum of black hole is of the form
\begin{equation}
\label{bek} A_n=\Delta A \cdot n,\quad n=0,1,2,...,
\end{equation}
where $\Delta A $ is the quantum of black hole area. Despite this
classical result there is still no general agreement on the precise
value of $\Delta A $; in the literature (see, for example,
\cite{med1} and references therein), the two alternative values are
mainly considered:
\begin{equation}
\label{val1} \Delta A =4\ln(k)l_P^{2},
\end{equation}
where $k$ is an positive integer, and
\begin{equation}
\label{val2} \Delta A =8\pi l_P^{2}.
\end{equation}
The specific value of $\Delta A $ is important for a statistical
definition of black hole entropy. According to statistical mechanics
the entropy of an ordinary object is the logarithm of the number of
microstates accessible to it, $\Delta \Gamma$, that is,
\begin{equation}
\label{ent1} S = \ln \Delta \Gamma.
\end{equation}
Since we assume that the entropy of a black hole should also have
the form (\ref{ent1}), it follows from (\ref{ent0}) - (\ref{val2})
that the number of microstates accessible to a black hole is
\begin{equation}
  \Delta \Gamma =
  \left\{
  \begin{array}{cll}
    k^{n}, & \mbox{in the case where  $\Delta A =4\ln(k)l_P^{2}$}, \\
    \exp (2\pi
n), & \mbox{in the case where  $\Delta A =8\pi l_P^{2}$}.
  \end{array}
  \right.
\label{num1}
\end{equation}
The number of microstates is intrinsically an integer. The value
$\Delta A =4\ln(k)l_P^{2}$ is consistent with this condition, but
the value $\Delta A =8\pi l_P^{2}$ is not. Since the value $\Delta A
=8\pi l_P^{2}$, as is well known from the literature, is not
restricted only to the semiclassical regime, this inconsistency
seems to compound a problem. It is little discussed in the
literature. Medved \cite{med1} was the first to consider it. Medved
suggested that if the Bekenstein-Hawking entropy does not have the
strict statistical interpretation of the form (\ref{ent1}), then the
two values (\ref{val1}) and (\ref{val2}) can be of comparable merit.
In this case there is no a problem of $\Delta A =8\pi l_P^{2}$.

In this paper I suggest an alternative solution of the problem.
Namely I suggest that the black hole entropy is really associated
with the number of microstates but, in contrast to ordinary matter
(\ref{ent1}), without the logarithm, that is,
\begin{equation}
\label{ent2} S_{bh} = 2\pi\cdot \Delta \Gamma,
\end{equation}
where the number of microstates for a given area is
\begin{equation}
\label{num2} \Delta \Gamma \equiv n= \frac{A}{8\pi l_P^{2}}.
\end{equation}
As is well known, a number of other entropy calculations have also
been proposed to explain black hole statistical mechanics (see, for
example, \cite{all} and references therein). But they all use the
usual expression (\ref{ent1}) with the logarithm. The point is that
every such a "calculation" is not a calculation in the ordinary
sense, but rather a new \emph{definition} of the black hole entropy,
which only be made precise by referring to the (still missing)
quantum theory of gravity. Moreover, none is yet very convincing.

The organization of this paper is as follows. In Sec. II we begin
with nonadditive properties of black hole and define a
microcanonical distribution with allowance for nonadditivity. In
Sec. III we show that quantization of black hole is nothing but
Landau quantization. We calculate Landau levels of black hole and
their degeneracy. The effect of noncommuting coordinates is
discussed. The new definition of black hole entropy is proposed in
Sec. IV. There the nature of the degrees of freedom responsible for
black hole entropy is elucidated. The applications of the new
interpretation are also presented. Finally, in Sec. V we consider
the holographic principle and suggest an explanation for the area
scaling $S_{bh} \sim A$ in the case where the degrees of freedom
does not reside on the horizon but are distributed in a spatial
volume.

\section{Black holes and nonadditivity}
\subsection{Motivation}

We begin with definitions. The essential reason for taking the
logarithm in (\ref{ent1}) is to make the entropy an \emph{additive}
quantity, for the statistical independent systems. If we can
subdivide a system into $n$, for example, separate subsystems and
each subsystem has $k$ states available to it, then the statistical
independence of these subsystems signifies mathematically that the
number of states for the composite system is the product of the
number of states for the separate subsystems \cite{lan1}:
\begin{equation}
\label{num3} \Gamma = k\times k\times k\times \cdot\cdot\cdot
=k^{n}.
\end{equation}
Then the additive property of the entropy defined as $\log$ of the
number of states follows from (\ref{num3}) directly:
\begin{equation}
\label{ent3} S=\ln \Delta\Gamma = n \ln k,
\end{equation}
that is, the total entropy of the system is $n$ times the entropy of
a single subsystem. It is these properties that are essentially used
in deriving the value $\Delta A =4\ln(k)l_P^{2}$. There are several
ways to obtain $\Delta A =4\ln(k)l_P^{2}$. A typical assumption is
that the horizon surface consists of $n$ independent patches of area
$\sim l_P^{2}$ and every patch has $k$ states available to it. Then
the total number of states is $\Delta\Gamma=k^{n}$, which is the
same as (\ref{num3}). Now assuming the usual interpretation of the
black hole entropy, we obtain $S=\ln \Delta\Gamma = n \ln k$, which
is just (\ref{ent3}). On the other hand, the entropy of a black hole
is related to the area $A$ of its horizon by the Bekenstein-Hawking
formula (\ref{ent0}). A comparison of these two expressions just
gives $\Delta A =4\ln(k)l_P^{2}$. So it is not a surprise that
$\Delta A =4\ln(k)l_P^{2}$ satisfies the condition (\ref{num1}) on
the number of states to be an integer. On the contrary, $\Delta A
=8\pi l_P^{2}$ is sought without any initial assumptions regarding
statistical interpretation of the Bekenstein-Hawking entropy; it
follows from the periodicity of the Euclidean black hole solutions,
underlying the black hole thermodynamics (see, for example,
\cite{ro}). In this case, as will be shown below, a new statistical
interpretation of the Bekenstein-Hawking entropy follows from
$\Delta A =8\pi l_P^{2}$.

The above derivation of $\Delta A =4\ln(k)l_P^{2}$ as well as the
classical formula for the entropy itself rely on the additivity
properties of ordinary matter and, more fundamentally, on the very
possibility of describing a given system as made up of independent
subsystems. However the black holes are not conventional systems:
they constitute nonadditive thermodynamical systems (for the sake of
simplicity, we shall not make a distinction between nonadditive and
nonextensive properties of black holes). As is well known, the fact
that the gravitational energy is nonadditive appears already in
Newtonian gravity. In general relativity a local definition of mass
is not possible; the ADM and Komar definitions of mass express this
very clearly. Moreover the black hole entropy (\ref{ent0}) goes as
the square of mass $M^{2}$ in a sharp contrast with the additive
character of entropy in ordinary thermodynamics. As emphasized by
Kaburaki \cite{ka} and also Arcioni and Lozano-Tellechea \cite{arc},
one has to consider a single black hole as a whole system; any
discussion related to the possibility of dividing it into subsystems
or to the additivity property of the black hole entropy simply does
not take place. The statistical independence is a postulate in
ordinary statistical physics and many its general results just fail
if this property is not assumed. This depart from the conventional
systems is closely related to the long-range behavior of
gravitational forces. Note that our proposal $\Delta\Gamma= n$ also
gives $S\propto n$ as in the case of conventional systems
(\ref{ent3}). But our proposal is inconsistent with any hypothesis
of the statistical independence. We know that if the number of
states for a compound system is a product of factors, each of which
depends only on quantities describing one part of the system, then
the parts concerned are statistically independent, and each factor
is proportional to the the number of states of the corresponding
part. In our approach the number $n$ can not be represented as such
a product. On the other hand, if the black hole constituents were
statistically independent, as in deriving $\Delta A
=4\ln(k)l_P^{2}$, the entropy (\ref{ent2}) would be nonadditive.

It is obvious that the above aspect of nonadditivity can not be
ignored in deriving the black hole entropy. Although the study of
nonadditive thermodynamics has been worked out to some extent (see,
for example, \cite{ka} - \cite{nonad} and references therein), there
is no (with rare exception \cite{can}) a concrete statistical model
of the black hole entropy with allowance for nonadditivity. It is
clear: we do not yet have a satisfactory quantum theory of gravity
whose classical limit is general relativity. But our task is
facilitated by the fact that the black hole area is quantized just
with the quantum $\Delta A =8\pi l_P^{2}$. Thus we can suggest a
more concrete statistical interpretation of the black hole entropy.

\subsection{Microcanonical distribution for a black hole with allowance for nonadditivity}
To apply statistical mechanics to a black hole we should at first to
define the distribution function. In statistical mechanics all
properties of a system are encoded in its distribution function
\cite{lan1}. For a quantum system the distribution function $w_n$
determines the probability to find the system in a state with energy
$E_n$. The determination of this function is the fundamental problem
of statistical physics. The form of the function is usually
postulated; its justification lies in the agreement between results
derived from it and the thermodynamic properties of a system.

We begin with conventional systems. The standard determination of
the distribution function for them is given in detail by Landau and
Lifshitz \cite{lan1}. Following Landau and Lifshitz consider an
ordinary isolated system consisting of quasi-isolated subsystems in
thermal equilibrium. According to Liouville's theorem the
distribution function of isolated system is an integral of the
motion. Due to the statistical independence of subsystems and, as a
consequence, multiplicativity of their distribution functions, the
logarithm of the distribution function must be not merely an
integral of the motion, but an additive integral of the motion. It
can be shown that the statistical state of a system executing a
given motion depends only on its energy. Thus we can deduce that the
logarithm of the distribution function must be a linear function of
its energy of the form
\begin{equation}
\label{func1} \ln w^{(a)}_n = \alpha^{(a)} + \beta E^{(a)}_n,
\end{equation}
with constant coefficients $\alpha$ and $\beta$, of which $\alpha$
is the normalization constant and $\beta$ must be the same for all
subsystems in a given isolated system; the suffix "$a$" refers to
the subsystem $a$. Note that assuming another dependence $\ln w$ on
$E$ we may not obtain an additive function on the right side of
(\ref{func1}); for example, $E^{2}$ is already a nonadditive
function. Since the values of nonadditive integrals of the motion do
not affect the statistical properties of ordinary system, these
properties can be described by any function which depends only on
the values of the additive integrals of the motion and which
satisfies Liouville's theorem. The simplest such function is
\begin{equation}
\label{func2} dw = const \times \delta (E -E_0)\prod \limits_a
d\Gamma_{a},
\end{equation}
where the number of states of the whole system $d\Gamma$ is a
product $d\Gamma=\prod \limits_a d\Gamma_a$ of the numbers
$d\Gamma_a$ of the subsystems (such that the sum of the energies of
the subsystems lies in the interval of energy of the whole system
$dE$). It defines the probability of finding the system in any of
the $d\Gamma$ states. The factor $const$ is the normalization
constant, $\delta (E -E_0)$ the Dirac delta function. The
distribution (\ref{func2}) is called microcanonical. Note that
(\ref{func1}) is nothing but the canonical distribution if we
identify $\beta=-1/T$, $\alpha=F/T$, $F$ being the free energy and
$T$ the temperature of the system.

As is easily seen, the statistical independence and additivity play
a crucial role in deriving the distribution function for the
conventional systems. Now consider the black holes. Because of the
nonadditivity, the black holes can not be thought as made up of any
independent subsystems. Therefore, if we want to establish the
distribution function for the black holes, we should remove the
restrictions of the statistical independence and additivity of
integrals of the motion for the subsystems of black hole. The
presence of the logarithm in (\ref{func1}) was just required by the
statistical independence of the subsystems. So dropping the
logarithm and suffix "a" in (\ref{func1}) we obtain
\begin{equation}
\label{func3} w_n = f(E_n),
\end{equation}
where $f(E_n)$ represents a nonadditive integral of the motion and
is a nonlinear function of the black hole energy. Besides the
energy, in an isolated classical system there is another integral of
motion - the phase volume occupied by the system $\Delta\Gamma$
(Liouville's theorem). It follows that any function of
$\Delta\Gamma$, in particular the entropy is also an integral of
motion \cite{sec}. Since the Bekenstein-Hawking entropy is
proportional to $M^{2}$ and nonadditive, it is reasonably assumed
that so is the integral of motion. Thus the simplest function
$f(E_n)$ compatible with this assumption is the square of energy, so
we can write
\begin{equation}
\label{func4} w_n = \gamma E_n^{2},
\end{equation}
where $\gamma$ is a constant coefficient. Note that here, as in
ordinary statistics, the form of the distribution function must be
regarded only as a postulate, to be justified solely on the basis of
agreement of its predictions with the thermodynamical properties of
black holes. Our considerations are intended to make it plausible,
and nothing more. As a result, the (canonical) distribution for a
subsystem (\ref{func1}) transforms to the (microcanonical)
distribution for the whole system. Similarly dropping the product
and suffix "a" in (\ref{func2}), we obtain
\begin{equation}
\label{func5} dw = const \times \delta (E -E_0)d\Gamma,
\end{equation}
or after integration,
\begin{equation}
\label{func6} w = const \times \Delta\Gamma,
\end{equation}
where $\Delta \Gamma$ is the number of states that accessible to the
whole system in a given state. The functions (\ref{func4}) and
(\ref{func6}) are obviously the same and satisfy the same
normalization condition $\sum \limits_n w_n =1$. So comparing
(\ref{func4}) and (\ref{func6}) we get
\begin{equation}
\label{func7} \Delta\Gamma \propto M^{2},
\end{equation}
where the energy of black hole is identified with its mass, $M$.
Thus we arrive at the conclusion that the statistical weight of
black hole should be proportional to the area
\begin{equation}
\label{func8} \Delta\Gamma =\frac{A}{\xi l^{2}_P},
\end{equation}
as is evident from dimensionality considerations. Here $\xi$ is a
new constant coefficient, $\xi > 1$, which cannot be defined exactly
from general considerations without assuming some dynamical model
for a black hole.

\section{Quantization of black hole as Landau quantization}

\subsection{Motivation}

As appears from the above, one has to consider a black hole as an
indivisible fundamental object, for example as an elementary
particle (in agreement with old idea of 't Hooft). But the
degeneracy factor of an elementary particle is relatively small.
Where does the large $\Delta\Gamma$ of black hole come from? Landau
quantization is interesting for the statistical interpretation of
black hole entropy mainly because of the macroscopic degeneracy of
Landau levels. This large degeneracy follows from the fact that the
orbital angular momentum $L_z$ of elementary particle can be
macroscopically large, proportional to the area of a sample. As will
be shown later, a black hole really has an intrinsic angular
momentum with such a property and the energy levels of black hole
are nothing but the Landau levels. Finally, Landau quantization is
important for black hole physics due to quantization of area and
noncommutating coordinates.

\subsection{Electron in two dimensions in a magnetic field}

Before we start out discussing Landau quantization of black hole we
need to define the conventional Landau quantization proper. For the
convenience of the reader we repeat the relevant material from
\cite{gas} without proofs, thus making our exposition
self-contained. As is well known from quantum mechanics, a magnetic
field quantizes the energy of an electron confined in two
dimensions. This is the basis of the conventional Landau
quantization. So we restrict our attention to the motion of a single
spinless electron confined to the $x-y$ plane in a perpendicular
magnetic field $\vec{B}=B\vec{z}$. In classical mechanics, the
centrifugal force is balanced by  the Lorentz force
\begin{equation}
\label{cycl1} \frac{m_ev^{2}}{r}=\frac{e}{c}vB,
\end{equation}
where all quantities have the standard meaning, so that a magnetic
field $\vec{B}$ forces an electron to move on a circular orbit at
the cyclotron frequency in the $x-y$ plane
\begin{equation}
\label{cycl2} \omega_c=\frac{v}{R_c},
\end{equation}
where $R_c$ is the cyclotron radius, $R_c=\sqrt{2m_eE_k}/(eB)$, and
$E_k$ is the kinetic energy of electron. For completeness we also
define the Larmor frequency
\begin{equation}
\label{cycl3} \omega_L=\frac{v}{2R_c},
\end{equation}
Next, introducing the angular momentum $L_z=mvr$ we obtain from
(\ref{cycl1})
\begin{equation}
\label{cycl4} \frac{m_ev^{2}}{2}=\omega_L L_z.
\end{equation}
In a quantum mechanical treatment, $L_z$ can take only discrete
values $m\hbar$, so that
\begin{equation}
\label{cycl5} \frac{m_ev^{2}}{2}=\omega_L \hbar m.
\end{equation}
The mean potential energy is just as large as the mean kinetic
energy, and the one-particle energy simply follows as the sum of
both:
\begin{equation}
\label{cycl6} E=\omega_c \hbar m.
\end{equation}
In the exact calculation, however, the zero-point energy also
appears. The restriction to positive components, $L_z>0$, is a
result of the chirality built into the problem by the magnetic field
(as will be shown later, in the black hole case this property is
caused by the Euclideanization of the black hole metric). The energy
levels (\ref{cycl5}) are degenerate; it appears that the degeneracy
is proportional to the area of the system. The macroscopically large
degeneracy corresponds to the fact that the center of a classical
circular orbit can be located anywhere in the $x-y$ plane.

In quantum mechanics, the Hamiltonian describing the cyclotron
motion of a single electron is
\begin{equation}
\label{ham1}
H=\frac{1}{2m}\left(\vec{p}+\frac{e}{c}\vec{A}\right)^{2}.
\end{equation}
Here $\vec{p}=(p_x, p_y)$ is the momentum operator and
$\vec{A}(x,y)$ is the vector potential. For the "symmetric gauge",
$\vec{A}=B(-y,x)/2$, the Hamiltonian (\ref{ham1}) can be written as
\begin{equation}
\label{ham2}
H=\frac{p_x^{2}}{2m_e}+\frac{m_e\omega_L^{2}x^{2}}{2}+\frac{p_y^{2}}{2m_e}+\frac{m_e\omega_L^{2}y^{2}}{2}+
\omega_L L_z.
\end{equation}
Note that the first two terms in $H$ form the Hamiltonian of an
isotropic two-dimensional oscillator. In the polar coordinates
defined by $x=r\cos\varphi$ and $y=r\sin\varphi$ the Hamiltonian
(\ref{ham2}) reads
\begin{equation}
\label{ham3}
H=-\frac{\hbar^{2}}{2m_e}\left[\frac{\partial^{2}}{\partial
r^{2}}+\frac{1}{r}\frac{\partial}{\partial
r}+\frac{1}{r^{2}}\frac{\partial^{2}}{\partial\varphi^{2}}\right
]+\frac{m_e\omega_L^{2}r^{2}}{2}-\omega_L i\hbar
\frac{\partial}{\partial\varphi}.
\end{equation}
It has eigenvalues
\begin{equation}
\label{en1} E=\hbar \omega_L(2n_r+1+m+|m|),
\end{equation}
where $n_r$ is the radial quantum number, $n_r=0,1,2,...$, and $m$
is the angular momentum quantum number, $m=0,\pm1, \pm2,...$ . The
energy levels are labeled by the principal quantum number $n$,
$n=n_r+(m+|m|)/2$, so that
\begin{equation}
\label{en2} E=\hbar \omega_c\left(n+\frac{1}{2}\right),\quad
n=0,1,2,...\, .
\end{equation}
These levels are called Landau levels. The lowest energy level has
$n_r=0$, $m=0,-1,-2,....$ and energy $E=\hbar \omega_c/2$. The first
excited level has $n_r=1$ and $m=0,-1,-2,...$, or $n_r=0$ and $m=1$,
etc. The eigenfunctions can be expressed in terms of the associated
Laguerre polynomials. Relatively simple are the eigenfunctions for
the lowest Landau level,
\begin{equation}
\label{fun1} \psi_{0,m}=\frac{1}{\sqrt{2\pi
|m|!}l_0}\left(\frac{r^{2}}{2l_0^{2}}\right)^{|m|/2}e^{im\varphi}e^{-r^{2}/4l_0^{2}},
\end{equation}
where $l_0$ is the characteristic length of the theory, the so
called magnetic length,
\begin{equation}
\label{leng1} l_0= \sqrt{\frac{\hbar}{(m_e\omega_c)}}\,.
\end{equation}
It follows that
\begin{equation}
\label{fun2} \langle
\psi_{0,m}|r^{2}|\psi_{0,m}\rangle=2(m+1)l_0^{2}.
\end{equation}
and
\begin{equation}
\label{fun3} \langle \psi_{0,m}|L_z|\psi_{0,m}\rangle=m.
\end{equation}
For increasing $m$ the wavefunction is localized along circles of
larger and larger radii. The degree of degeneracy can be determined
from the requirement that the radius for the largest $m$ should be
inside our system, for example, a disk of radius $R$,
\begin{equation}
\label{deg1} 2l_0^{2}(m+1)=R^{2}.
\end{equation}
Expressed in terms of the area $A=\pi R^{2}$, this gives
\begin{equation}
\label{deg2} m+1=\frac{A}{2\pi l_0^{2}}.
\end{equation}
This is also true for higher Landau levels. We now return to the
expression for the energy, (\ref{en1}). Because of the smallness of
$\hbar$, the energy can only be of macroscopic magnitude for
reasonable $B$, if $(2n_r+1+m+|m|)$ is very large. So we have two
cases: (i) $m<0$, and (ii) $m>0$. It appears that in the case (i)
$n_r$ is large so the wavefunctions do not satisfy some natural
conditions. There is no such a problem in the case (ii). If $m>0$,
the factor is $(2n_r+1+2m)$, and it can be large, with $n_r$ small,
provided that $m$ is large. The energy now is
\begin{equation}
\label{en3} E=\omega_c \hbar m,
\end{equation}
in agreement with the classical result (\ref{cycl6}). Note that
$L_z$ is positive, as expected. Note also that for large $m$ the
degree of degeneracy is
\begin{equation}
\label{deg3} m=\frac{A}{2\pi l_0^{2}}.
\end{equation}
This is nothing but quantization of the area of electron orbit.

\subsection{Black hole in two dimensional Euclidean space}

It appears that quantization of a black hole is nothing but the
Landau quantization. The matter is that kinematics of a black hole
in two dimensional Euclidean Rindler space is similar to that of an
electron in two dimensions in a magnetic field. So we start with
Rindler space. It is well established \cite{all}, that in the
near-horizon approximation the metric of an arbitrary black hole can
be reduced to the Rindler form. In this approximation the first law
of black hole thermodynamics for a Schwarzschild black hole takes
the form \cite{sus}
\begin{equation}
\label{rind1} dE_R=T_RdS_{bh},
\end{equation}
where $E_R$ is the Rindler energy, $E_R=2GM^{2}$, $T_R$ is the
Rindler temperature, $T_R=1/(2\pi)$, and $S_{bh}$ is the
Bekenstein-Hawking entropy. These quantities are related by
\begin{equation}
\label{rind2} E_R=T_RS_{bh}.
\end{equation}
In \cite{ro} quantization of black hole area (\ref{bek}) and value
$\Delta A =8\pi l_P^{2}$ were derived from the quantization of the
angular momentum associated with the Euclidean Rindler space of a
black hole. In transforming from Schwarzschild to Euclidean Rindler
coordinates the Schwarzschild metric becomes
\begin{equation}
\label{metr3}ds_E^{2}\approx(k\rho)^{2}dt^{2}+d\rho^{2}+\frac{1}{4k^{2}}d\Omega^{2},
\end{equation}
where $\rho$ is the proper distance from the horizon and the
constant $k$ coincides with the surface gravity of a Schwarzschild
black hole, $k = 1/4GM$. This metric is the product of the metric on
a two-sphere with radius $2GM$ (the last term) and the Euclidean
metric
\begin{equation}
\label{metr4}ds_E^{2}=\rho^{2}d(kt)^{2}+d\rho^{2}.
\end{equation}
The metric (\ref{metr4}) has a coordinate singularity at $\rho = 0$
(corresponding to the gravitational radius $R_g= 2GM$). Regularity
is obtained if $kt$ is interpreted as an angular coordinate with
periodicity $2\pi$
\begin{equation}
\label{om} \omega=kt=\frac{t}{4GM}.
\end{equation}
($t$ itself has then periodicity $8\pi GM$ which, when set equal to
$\hbar/T_H$, gives the Hawking temperature $T_H$). This periodicity
plays the same role in quantization of black hole as a magnetic
field in the conventional Landau quantization. On the other hand,
according to quantum mechanics the angle (\ref{om}) is conjugate to
the $z$th component of the angular momentum. Therefore, as was
suggested in \cite{ro}, the Rindler energy $E_R$ should be
reinterpreted as the $z$th component of an angular momentum operator
$i\hbar\partial/\partial\omega$ with eigenvalue
\begin{equation}
\label{angul} L_z= 2GM^{2}
\end{equation}
(that is why we use the different notations, $E_R$ and $L_z$, for
the same value $2GM^{2}$). Since $L_z=m\hbar$, $m=0,\pm1, \pm2,
...$, the value $2GM^{2}$ is now quantized. The negative integers
$m$ correspond to the region $r<R_g$. But the Euclidean Rindler
spacetime has no region corresponding to the region $r<R_g$ in the
Lorentzian spacetime, so the negative integers can be ruled out. In
\cite{ro}, quantization of $L_z$ was interpreted as quantization of
the black hole area,
\begin{equation}
\label{quant2} \frac{A}{8\pi l_p^{2}}=m,\quad m=0,1,2,...\,.
\end{equation}
In \cite{ro} it was shown that this conclusion is also valid for a
generic Kerr-Newman black hole. A refined version of this approach
extended to generic theories of gravity was presented by Medved
\cite{med2}. The angular momentum (\ref{angul}) can be also written
in the usual classical form
\begin{equation}
\label{quant3} L_z=Mvr
\end{equation}
and associated with some intrinsic motion if we identify $M$ with
the mass of a body which moves in a circle of the radius $r=R_g$
with the linear velocity $v\equiv c=1$. This does not mean however
that our system (i.e. a black hole) represents a rigid rotator,
rather, as will be shown below, it represents a harmonic oscillator.
Since a black hole as a two-sphere has circumference $2\pi R_g$ the
period of such a "motion" is $2\pi R_g$  and the angular frequency
is $1/R_g$; by analogy with (\ref{cycl2}) we shall call this
frequency the \emph{cyclotron frequency} and denoted by $\omega_c$,
\begin{equation}
\label{cyc1} \omega_c= \frac{1}{R_g}.
\end{equation}
Since the Rindler time $\omega$ is related to the Schwarzschild time
$\tau$ by (\ref{om}), a field quantum with Rindler frequency $\nu_R$
is seen by a distant Schwarzschild observer to have a red shifted
frequency $\nu=\nu_R/(4GM)$. From this it follows \cite{sus} that
the temperature as seen by the distant observer is just the Hawking
temperature $T_H=T_R\times1/(4GM)$. By analogy with (\ref{cycl3}) we
shall call the quantity $1/(4GM)$ the \emph{Larmor frequency} and
denote by $\omega_L$,
\begin{equation}
\label{cyc2} \omega_L= \frac{1}{2R_g}
\end{equation}
(it is just half of the cyclotron frequency, $\omega_L=\omega_c/2$).
We now return to the expression for the Rindler energy,
(\ref{rind2}). Taking into account (\ref{angul}) and (\ref{cyc2}),
the expression (\ref{rind2}) can be rewritten as
\begin{equation}
\label{temper2} \omega_L L_z=T_HS_{bh}
\end{equation}
(the entropy $S_{bh}$ is an invariant and is not red shifted). Since
$M=2T_HS_{bh}$ and $L_z=m\hbar$, we can write
\begin{equation}
\label{mas1} M=2\omega_L \hbar m,
\end{equation}
or equivalently
\begin{equation}
\label{mas2} M=\omega_c \hbar m.
\end{equation}

\subsection{Landau levels of black hole}

Continuing our analogy with Landau quantization, we may expect that
in more general quantum mechanical case the Hamiltonian of black
hole has a similar to (\ref{ham2}) form, except that now it is
defined in the two dimensional Euclidean plane (\ref{metr4}) and all
quantities relating to the electron are replaced by the
corresponding quantities relating to the black hole:
\begin{equation}
\label{ham4}
H=\frac{P_x^{2}}{2M_\ast}+\frac{M_\ast\omega_L^{2}x^{2}}{2}+\frac{P_y^{2}}{2M_\ast}+\frac{M_\ast\omega_L^{2}y^{2}}{2}+
\omega_L L_z,
\end{equation}
or in polar coordinates $\rho-\omega$,
\begin{equation}
\label{ham5}
H=-\frac{\hbar^{2}}{2M_\ast}\left[\frac{\partial^{2}}{\partial
\rho^{2}}+\frac{1}{\rho}\frac{\partial}{\partial
\rho}+\frac{1}{\rho^{2}}\frac{\partial^{2}}{\partial\omega^{2}}\right
]+\frac{M_\ast\omega_L^{2}\rho^{2}}{2}+\omega_L L_z.
\end{equation}
Here $\omega_L$ is the Larmor frequency of black hole (\ref{cyc2})
and $L_z$ is the angular momentum operator introduced near
(\ref{angul}). Since the total energy of electron in a magnetic
field is twice the kinetic energy $m_e/2$, we replace the electron
mass by $M_\ast=M/2$, $M$ being the mass of black hole. Accordingly,
the magnetic length of electron is replaced by the characteristic
length of black hole $l_\ast$:
\begin{equation}
\label{leng2} l_0 \rightarrow l_\ast= \sqrt{\frac{\hbar}{M_\ast
\omega_c}}.
\end{equation}
It is important to emphasize that in agreement with nonadditivity of
a black hole there are no particular $\Delta m_i$ in (\ref{ham4}),
only the total $M$. The Hamiltonian (\ref{ham4}) can be postulated
from the very beginning. From (\ref{ham4}) it follows that a black
hole is a two-dimensional isotropic oscillator with an additional
interaction $\omega_L L_z$, like electron in a magnetic field
(\ref{ham2}). Since in the black hole case $L_z\geq 0$, the
eigenvalues of the Hamiltonian are
\begin{equation}
\label{en3} E=\hbar \omega_L(2n_r+1+2m)
\end{equation}
where $n_r$ is the radial quantum number, $n_r=0,1,2,...$ and $m$ is
the angular momentum quantum number, $m=0,1,2,...$. Analogously to
(\ref{en2}),
\begin{equation}
\label{en4} E=\hbar \omega_c\left(n+\frac{1}{2}\right),\quad
n=0,1,2,... .
\end{equation}
where $n=n_r+m$ is the principal quantum number. By analogy with the
energy levels of an electron we call (\ref{en4}) the \emph{Landau
levels} of a black hole. The lowest level has $n_r=0$ and $m=0$. It
may seem strange that there is an energy $E_0=\hbar \omega_c/2$ in a
state with $m=0$ ($M=0$). But $\hbar\omega_c\rightarrow\infty$ when
$M=0$, so a state with the zero-point energy in the absence of a
black hole has no real physical meaning. The zero-point energy has
however a significance for the higher energy levels. The energy
difference between the subsequent Landau levels is  $\hbar\omega_c$.
This gap decreases with increasing $M$ and is equal in order of
magnitude to $T_H$. It will be discussed in detail later. In the
semiclassical limit $m\gg1$  we obtain from (\ref{en3})
\begin{equation}
\label{en6} E=\hbar \omega_c m,
\end{equation}
which is the same as (\ref{mas2}).

\subsection{Degree of degeneracy of Landau levels}

We know that a black hole has the entropy so that each Landau level
should be degenerate. But where does this degeneration come from? As
mentioned above, the complete Euclidean Schwarzschild space
(\ref{metr3}) has the structure $R^{2}\times S^{2}$. The matter is
that the energy of black hole does not depend on a point of the
two-sphere at which the Euclidean space (\ref{metr4}) can be
attached. Since we associate the Euclidean space with internal
cyclotron motion, we can also say that the degeneracy corresponds to
the fact that the center of the motion can be located anywhere on
the two-sphere (in other words, all axes of rotation are physically
equivalent). If the accuracy with which this point can be determined
coincides with the size of the area quantum $\Delta A = 8\pi
l_P^{2}$, then the degeneracy factor is given by $\Delta\Gamma=
A/(8\pi l_P^{2})$. This nothing but the angular momentum number
(\ref{quant2}). As is well known, the energy levels of a system
whose angular momentum is conserved are always degenerate. It is
clear that $L_z$ is conserved. Since in the black hole case $L_z$
can take only positive values and zero,
\begin{equation}
\label{deg4} \Delta\Gamma = m + 1.
\end{equation}
For a typical black hole $m\gg1$ so
\begin{equation}
\label{deg5} \Delta\Gamma = m.
\end{equation}
On the other hand, $L_z$ is associated with a rotation in the
Euclidean Rindler space through an angle $\omega$. In the
semiclassical description for any rotational degrees of freedom the
number of accessible states equals the total accessible phase-space
volume divided by the volume of one state, $2\pi\hbar$:
\begin{equation}
\label{num4} \Delta\Gamma=\frac{\int d\omega dL_z}{2\pi\hbar}.
\end{equation}
Taking into account (\ref{quant2}) and the fact that the angular
orientation is unconstrained, so that the integral over $d\omega$
gives $2\pi$, we again obtain $\Delta\Gamma=m$. Note that the black
hole is degenerate with respect to $L_z$ exactly as the electron in
a magnetic field. So we can determine the degree of degeneracy of
black hole from the corresponding formulas for the electron
replacing all quantities relating to the electron by the
corresponding quantities related to the black hole. From
(\ref{leng2}) it follows that for the black hole the characteristic
length is $l_\ast=2l_P$. Substituting this value in (\ref{deg3})
instead of $l_0$, we obtain $\Delta\Gamma=A/(8\pi l_P^{2})$, as
expected.

\subsection{Noncommutative geometry}

Coordinate noncommutativity is one of the most fascinating effects
of the Landau quantization. It appears \cite{sh}, that in the limit
of very large magnetic field $B$ the energy difference between the
subsequent Landau levels $\hbar\omega_c\rightarrow \infty$ so that
an electron is restricted to the lowest Landau level. As a result,
the two coordinates of the $x-y$  plane obey the same commutation
relations as the momentum $p_x$ and the position $x$ in quantum
mechanics:
\begin{equation}
\label{non1} [x,y]=il_0^{2}.
\end{equation}
Thus, the two dimensional coordinate space becomes the \emph{phase
space} for the system. As mentioned above, the area of one state in
phase space is $\triangle p_x\triangle x=2\pi \hbar$, so that
\begin{equation}
\label{non2} \Delta x \Delta y=2\pi l_0^{2}
\end{equation}
(here $l_0^{2}$ plays a role of $\hbar$). Therefore the physical
plane $x-y$ can be thought of as divided into the patches of area
$2\pi l_0^{2}$ where the center of the motion can be localized. Note
that the phenomenon of noncommuting plane is not specific to the
lowest Landau level but can be obtained by projecting to an
arbitrary finite number of Landau levels \cite{mac}. Since the large
$B$ limit corresponds to small $m_e$, we can obtain a similar
relation for a black hole setting $M \rightarrow 0$ and replacing
$l_0$ by $l_\ast$ in the Lagrangian of black hole. As a result, we
get
\begin{equation}
\label{non2} \Delta x \Delta y=8\pi l_P^{2}.
\end{equation}
We can also obtain a similar relation for the space-time
coordinates. Since $L_z$ is conjugate to the angle (\ref{om}), we
have
\begin{equation}
\label{non3}[L_z,\frac{t}{4GM}]=i\hbar
\end{equation}
or
\begin{equation}
\label{non4}[R_g,t]=i\hbar4G.
\end{equation}
From this it follows that
\begin{equation}
\label{non5} \Delta r \Delta t=8\pi l_P^{2},
\end{equation}
as required.

\section{The black hole entropy}

\subsection{Definition}

Now we can define the entropy of black hole. The entropy plays a
particularly fundamental role when the microcanonical ensemble is
used. According to the standard formula for the entropy we would
have to take the logarithm of $\Delta\Gamma$. But in this case the
generalized second law of black hole thermodynamics would be
violated. The argument involves a well-known example with the
collision of black holes: two identical black holes collide, merge,
radiating gravitational wave energy, and form a third black hole.
According to (\ref{ent1}), the initial entropy of the system is
\begin{equation}
\label{ent6} S_i=2\ln \Delta\Gamma_i= 2\ln \left(\frac{A}{8\pi
l_P^{2}}\right).
\end{equation}
On the one hand, the final entropy is bounded from above by
\begin{equation}
\label{ent7} S_f=\ln \Delta\Gamma_f= \ln \left(\frac{4A}{8\pi
l_P^{2}}\right).
\end{equation}
On the other hand, by virtue of the generalized second law it must
be greater then initial entropy. So we have
\begin{equation}
\label{ent8} 2\ln \left(\frac{A}{8\pi l_P^{2}}\right) < S < \ln
\left(\frac{4A}{8\pi l_P^{2}}\right).
\end{equation}
As is easily seen, these inequalities are satisfied only for $A <
32\pi l_P^{2}$. This means that the standard interpretation of the
entropy in terms of the logarithm of $\Delta\Gamma$ violates the
generalized second law. Moreover, since $m$ takes not only positive
integral values but also zero, the entropy as the logarithm of
$\Delta\Gamma$ makes no sense at all. Thus we conclude that the
statistical interpretation of the Bekenstein-Hawking entropy is true
only if $\log$ is deleted from the Boltzmann formula (\ref{ent1}),
that is,
\begin{equation}
\label{ent9} S_{bh}=2\pi \cdot \Delta\Gamma = 2\pi m.
\end{equation}
This is nothing but the angular momentum quantization condition on
the phase of wavefunction: if the eigenfunction of $L_z$ is to be
single-valued, it must be periodic in phase, with period $2\pi$.
Note that factor $2\pi$ was already noticed in the literature in a
topological context \cite{teit}. In particular Bunster (Teitelboim)
and Carlip noted that the overall factor in front of the area,
usually quoted as one fourth in units where Newton's constant is
unity, is really the Euler class of the two-dimensional disk. A
number of proposals were proposed to quantize the entropy. Prominent
among others, besides the classical works of Bekenstein \cite{bek},
are those of Barvinsky and Kunstatter \cite{bar}, Padmanabhan and
Patel \cite{paddy1}, Romero, Santiago and  Vergara \cite{rom}, and
also Dolan \cite{dol}. It is important to notice here the following.
Although all these researchers obtained the required spectrum
$S_{bh}=2\pi m$, there is an important difference between their
result and ours: in their spectrum $m$ is simply a non-negative
integer, in ours it is the statistical weight of black hole,
$\Delta\Gamma = m$.

\subsection{The nature of the degrees of freedom}

As appears from the above, a Schwarzschild black hole is completely
described (at least in the semiclassical approximation) by one
quantum number - the angular momentum number $m$. So, by definition,
the black hole has \emph{one} degree of freedom. At first sight it
may seem that the horizon surface splits into $m$ elementary patches
of area $\Delta A= 8\pi l_P^{2}$. This is not the case; the number
$m$ does not mean that horizon is really divided into $m$ elementary
figures with specific shape and localization as a globe with
quadrangles formed by parallels of latitude and meridians of
longitude. According to nonadditivity of black hole, a black hole
cannot be thought as made up of any independent constituents; the
black hole is an indivisible fundamental object, like the electron.
On the other hand, although the black hole energy, like the energy
of the electron motion, is the sum of $m$ quanta with energy
$\hbar\omega_c$, this does not mean that a black hole (or an
electron) consists of $m$ "photons". Thus the number $m$ is not the
number of black hole constituents. Instead, it is \emph{the number
of distinguishable ways} to distribute a patch of area $8\pi
l_P^{2}$ over the horizon. This is its physical meaning. But the
number $m$ can have more deeper nature. The questions then arising,
however, have as yet hardly been studied at all.

\subsection{Applications: mean separation between energy levels of
black hole and Poincar\'{e} recurrence time}

The most distinctive feature of our interpretation of black hole
entropy is that the statistical weight $\Delta\Gamma \sim S_{BH}$ in
contrast to $\Delta\Gamma \sim \exp (S_{BH})$ of the usual
interpretation.  Here we consider the cases where the difference
between the old and new interpretations of black hole entropy can
manifest itself most clearly. We begin with energy spectrum.
According to (\ref{en4}) the energy spacing between the subsequent
Landau levels is
\begin{equation}
\label{mass1} \Delta E=\hbar\omega_c.
\end{equation}
This agrees with the characteristic value of Hawking radiation $\sim
T_H$. This value however does not agree with estimation obtained
from the usual definition of entropy. The entropy of an ordinary
system (\ref{ent1}), by definition, is the logarithm of the number
of states $\Delta\Gamma$ with energy between $E$ and $E+\delta{E}$.
The width $\delta{E}$ is some energy interval characteristic of the
limitation in our ability to specify absolutely precisely the energy
of a macroscopic system. Dividing $\delta{E}$ by the number of
states $\exp(S)$ we obtain the mean separation between energy levels
of the system \cite{lan1}:
\begin{equation}
\label{main1}\langle \Delta{E} \rangle = \delta{E} \exp(-S).
\end{equation}
The interval $\delta{E}$ is equal in order of magnitude to the mean
canonical-ensemble fluctuation of energy of a system. However a
Schwarzschild black hole has the negative specific heat $C_{v}$,
$C_{v} = -8\pi G M^{2}$, so that the energy fluctuations calculated
in the canonical ensemble have formally negative variance:
$\langle(\delta{E})^{2}\rangle=C_{v}T_H^{2} \sim - m_P^{2}$, where
$T_H$ is the Hawking temperature. The situation is quite different
if a black hole is placed in a reservoir of radiation and the total
energy of the system is fixed \cite{haw}. In this case a stable
equilibrium configuration can exist if the radiation and black hole
temperatures coincide, $T_{rad}=T_H\equiv T$, and $E_{rad}< M/4$,
where $E_{rad}$ is the energy of radiation . The latter condition
can be reformulated as the restriction on the volume of reservoir
$V$, $4aVT^{5}< 1$, where $a$ is the radiation constant . According
to this condition Pavon and Rub\"{i} found \cite{pav} that the mean
square fluctuations of the black hole energy (mass) is given by
\begin{equation}
\label{delta1} \langle (\delta E)^{2}\rangle =(1/8\pi) T^{2} Z,
\end{equation}
$Z$ being the quantity $4aVT^{3}/(1-4aVT^{5})$, $G=c=\hbar =1$ and
the Boltzmann constant $k_B=(8\pi)^{-1}$. It is clear that
(restoring $G$, $c$, $\hbar$, and $k_B$)
\begin{equation}
\label{delta2} \langle (\delta E)^{2}\rangle \sim
\frac{m_P^{4}}{M^{2}}
\end{equation}
and
\begin{equation}
\label{delta3} \langle (\delta E)\rangle \sim \frac{m_P^{2}}{M}.
\end{equation}
In the quantum mechanical description, the accuracy with which the
energy of a black hole can be defined by a distant observer is
limited by the time-energy uncertainty relation as well as by the
decrease of the mass of the black hole due to transition from a
higher energy level to a lower one. The lifetime of a state $E_n$ is
proportional to the inverse of the imaginary part of the effective
action \cite{muk}; less formally, it is the time needed to emit a
single Hawking quantum and this is proportional to the gravitational
radius $R_g$. So $\delta{E}_q \sim 1/R_g$, where I have added the
subscript "q" to refer to the quantum uncertainty. This is just the
zero-point energy of black hole $E_0=\hbar\omega_c/2$ mentioned
above. On the other hand, $\delta{E}_q\sim T_H$ due to transition
from the state $n$ to the state $n-1$. As is easily seen,
$\delta{E}_q$ is of the same order of magnitude as (\ref{delta3}).
So we obtain the mean separation between energy levels for a black
hole
\begin{equation}
\label{main2}\langle \Delta{E} \rangle \sim  \frac{m_P^{2}}{M}
\exp(-S_{bh}).
\end{equation}
This value is however exponentially smaller than (\ref{mass1}). Thus
a problem arises. It has not yet received attention in the
literature. The point is that the energy interval $\delta {E}$
contains only a single state. But in this case statistics is not
applicable; by definition the statistical treatment is possible only
if $\delta{E}$ contains many quantum states. This means that the
formula (\ref{main2}) is not applicable to a black hole (contrary to
what is assumed in many works \cite{sred}). Nevertheless we may
attempt to define $\langle \Delta{E} \rangle$ with the help of the
new interpretation of black hole entropy. Namely, taking into
account the relation $S_{BH}\sim \Delta\Gamma = m$ we get
\begin{equation}
\label{main4} \langle \Delta{E} \rangle \sim \frac{\Delta E}{\Delta
m}\sim \frac{dM}{dS_{bh}}=T_H;
\end{equation}
this agrees with (\ref{mass1}), as it should. Thus the first law of
thermodynamics defines the energy spacing of a black hole.

One further remark is required concerning the equipartition theorem;
it is closely connected with energy spacing of a black hole.
Recently Padmanabhan \cite{paddy2} and Verlinde \cite{ver} set up
very interesting hypotheses concerning the nature of gravity. One of
the crucial ingredients of their hypotheses is the claim that the
horizon degrees of freedom/bits satisfy the equipartition theorem.
As is well known \cite{lan1}, the theorem is valid only if the
thermal energy $k_BT_H$ is considerably larger than the spacing
between energy levels of a system. As is easily seen, if the energy
spacing of black hole levels were exponentially small in the black
hole entropy as (\ref{main2}), the theorem would be valid, since in
this case $k_BT_H \gg \Delta E$. But in fact the energy spacing of
black hole levels is the same order of magnitude as $k_BT_H$,
$\Delta E \sim k_BT_H $. Therefore the equipartition theorem is not
valid for the black holes. On the other hand, we have the formula
(\ref{en6}) which seems to support the assumption of Padmanabhan and
Verlinde, if we identify $m$ with the number of "bits". How can that
be? The point is that a black hole has no any independent
constituents or "bits". That is why the classical equipartition
theorem is not applied to a black hole. The number $m$ in
(\ref{en6}) is not the number of "bits" but the number of ways to
distribute a patch of area $8\pi l_P^{2}$ over the horizon. Thus
Padmanabhan and Verlinde are right only in a sense that we can
extract the frequency $\omega_c$ (rather than the temperature) from
(\ref{en6}).

As mentioned above, a description of a black hole via a thermal
ensemble is inappropriate. From the point of view of statistics this
could be explained as follows. According to the canonical
distribution, the probability $p_i$ of a system being in a state of
energy $E_i$ is proportional to the Boltzmann factor,
\begin{equation}
\label{canon1} p_i \sim \Delta \Gamma_i
\exp\left(-\frac{E_i}{T}\right),
\end{equation}
where $T$ is the temperature of the system, $T=T_{heat\; bath}$.
Assuming the usual interpretation of entropy (\ref{ent1}), the
statistical weight of black hole should grow with $M$ as
\begin{equation}
\label{func9} \exp (4\pi GM^{2}).
\end{equation}
In that case, the total probability diverges. However, in the case
where the entropy is given by (\ref{ent2}), the statistical weight
grows as
\begin{equation}
\label{func10} 2GM^{2},
\end{equation}
and the probability can converge. So, it seems that the usual value
of the statistical weight (\ref{func9}) better explains the
breakdown of the canonical ensemble for black holes than suggested
(\ref{func10}). But this is not the case. First, it is clear that an
indefinitely large heat bath is gravitationally unstable. On the
other hand, there is always a size of bath at which the interaction
energy between the members of ensemble is not negligible. In both
cases the the canonical distribution is not applicable. Secondly, a
black hole possesses a very special property which singles it out;
namely, its size and temperature are not independent parameters. As
a result, the temperature of black hole does not remain constant at
the different $E_i$ so that $T_H \neq T_{heat\; bath}$, contrary to
the definition of canonical ensemble. This is irrespective to the
form of the statistical weight. Note that the ordinary
self-gravitating systems, for example, stars and galaxies, also have
negative heat capacity. And although their statistical weights grow
not so fast as (\ref{func9}), they can not be in a thermal
equilibrium with a heat bath \cite{lind}. Thus the apparent
divergence of canonical distribution, in the case where the
statistical weight grows as (\ref{func9}), cannot be an evidence in
support of the usual interpretation of black hole entropy.

One more case, where the old and new interpretations of black hole
entropy give different answers - the Poincar\'{e} recurrence time.
According to the Poincar\'{e} recurrence theorem \cite{huang}, any
state of an isolated finite system continuously returns arbitrarily
close to its initial value in a finite amount of time (the
Poincar\'{e} recurrence time, $t_{r}$). For an ordinary system this
time is exponentially large in the thermodynamical entropy of the
system:
\begin{equation}
\label{time1}t_{r}\sim t_0 \exp (S),
\end{equation}
where $t_0$ is the time required for a fluctuation, once it occurs,
to again degrade. To apply the theorem to the black holes one has to
place a black hole in a finite reservoir with a fixed total energy.
We shall assume that all requirements of the Poincar\'{e} recurrence
theorem are satisfied and a black hole is in equilibrium with its
own radiation in an appropriate reservoir. We also assume that the
total entropy is dominated by the entropy of a single black hole. In
this case the Poincar\'{e} recurrence time for a black hole is given
by
\begin{equation}
\label{time2}t_{r}\sim t_0 S_{bh}.
\end{equation}
Then assuming that for a black hole $t_0$ is not smaller than its
lifetime, $\sim t_P(M/m_P)^{3}$, we obtain
\begin{equation}
\label{time3}t_{r}\sim t_P \left(\frac{M}{m_P}\right)^{5}.
\end{equation}
This time is considerably smaller than (\ref{time1}). There is
however an obvious explanation for this behavior: due to the
long-range attractive character of gravitational forces matter is
very unstable with respect to the clumping.

\section{The nature of the area scaling $S\propto A$}

This Section has more speculative character and is not related with
previous one directly. Early in counting the number of degrees of
freedom responsible for black hole entropy we assumed that they all
reside on the horizon. In this case the area scaling $S\propto A$ is
natural. There is however a hypothesis that the degrees of freedom
are distributed in a spatial volume. But in this case the most of
them are not involved in black hole thermodynamics. An ordinary
quantum field theory underlying thermodynamics can not explain this
fact. In \cite{hooft} it is noted that in a fundamental fermion
theory with a cut-off at the Planck scale the total number of
independent quantum states in a given volume $V$ of space is $\Delta
\Gamma \sim k^V$, where $k$ is the number of state of a single
fermion. Note that if there are any fundamental bosons, the number
of possible states should be infinite.  So the entropy is
proportional to the volume instead of being proportional to the
area. An explanation is that most of the states of field theory are
not observable since their energy is so large that they are confined
inside their own gravitational radius. In this way gravitation
reduces the number of physical degrees of freedom so that the number
of states grows exponentially with the area instead of the volume.
It is then conjectured that all degrees of freedom are resided on
the surface of volume. This is called the holographic hypothesis.

We want to suggest an alternative mechanism for the area scaling
$S\propto A$. Our proposal is based on the analogy with electrons in
metal \cite{somm}. As is well known, according to the principle of
equipartition of energy, the conduction electrons in a metal viewed
as a classical electron gas should make a contribution $3N/2$ ($N$
is the number of electrons) to the heat capacity of the metal. In
reality the electronic contribution to the heat capacity at room
temperature is only of the order of one per cent of the classical
value. This means that only small fraction of electrons participates
in thermal equilibrium, not the total number of free electrons. The
observation is completely unexplained by classical theory, but is in
good agreement with quantum statistics. It turns out that the
difficulty disappears if it is taken into account that electron gas
possesses the properties of a highly degenerate Fermi gas.  We now
proceed to explain the area scaling $S\propto A$ basing our
considerations on the analogy with electrons in metal. First we
assume that at the very fundamental level of matter there exist the
fundamental fermions. Then suppose that $N$ fundamental fermions
with spin $1/2$ are uniformly distributed in a spatial region of
volume $\sim R_g^{3}$ with spacing $\sim l_P$, so that $N\sim
(R_g/l_P)^{3}$. At $T = 0$ the first $N/2$ states up to the energy
$E_{max}$ will be "completely" filled, with two fermiones with
opposite spins per state (in accordance with the Pauli principle),
while all states with $E> E_{max}$ will be empty (the limiting
energy $E_{max}$ is generally referred to as the Fermi energy of the
system and is denoted by the symbol $\varepsilon_F$). It is obvious
that there is one and only one way of achieving this arrangement
with indistinguishable particles. Therefore $S = 0$. Remember that
according to Dirac's pre-quantum field theory picture the vacuum
consists of states of positive and negative energies, with the
negative energy states completely filled and states of positive
energy empty. Note that this picture applies only to fermions. In
spirit of Dirac's picture we assume that all our fundamental
fermions are also unobservable (perhaps they have negative energies
due to the effects of gravitation). Suppose now that the energy
levels are uniformly distributed so that the energy difference
between neighboring levels is $\Delta E=2 \varepsilon_F/N$. If one
goes from the temperature $T = 0$ to a temperature $T > 0$ slightly
above zero, then some of the fermions will be thermally excited from
states just below the Fermi energy to states just above the Fermi
energy. Then the number of fermions close to the Fermi surface which
increase their energy by an amount $\sim k_B T$ is given
approximately by
\begin{equation}
\label{num7} \Delta N \sim \frac{k_B T}{\Delta E}\sim N\frac{k_B
T}{\varepsilon_F}.
\end{equation}
Since the temperature of a black hole  $T \sim 1/R_g$, we have
\begin{equation}
\label{num8} \Delta N \sim \left(\frac{R_g}{l_P}\right) ^{3}
\frac{1}{\varepsilon_F R_g}\sim (\varepsilon_F^{-1}l_P^{-3})
R_g^{2}\propto A.
\end{equation}
This means that the volume remains uninfluenced by the rise in
temperature. That is why the area, not volume, is relevant in black
hole thermodynamics. Since each of the excited fermions receives an
additional energy of $\sim k_B T$, the internal energy of black hole
will be
\begin{equation}
\label{en} E\sim k_B T \times\Delta N \sim
(\varepsilon_F^{-1}l_P^{-3}) R_g.
\end{equation}
This is just the black hole mass if we identify $G^{-1}\sim
\varepsilon_F^{-1}l_P^{-3}$ or $\varepsilon_F\sim m_P$. So the heat
capacity, $C_V$, is given by
\begin{equation}
\label{heat} C_V = \left(\frac{\partial E}{\partial T}\right)_V \sim
-(\varepsilon_F^{-1}l_P^{-3}) R_g^{2}.
\end{equation}
If we now deduce the entropy from the heat capacity, $C_V=T(\partial
S/\partial T)$, we get
\begin{equation}
\label{ent10} S =|C_V| \sim \Delta N \propto A,
\end{equation}
which is what had to be proved.


\begin{thebibliography}{99}
\bibitem{sec}
Of course, in a practical sense the entropy is not conserved. This
is because the region $\Delta\Gamma$  becomes very complicated like
a fractal, and if we coarse grain the phase space, it will appear
that $\Delta\Gamma$ is growing. This is the origin of the second law
of thermodynamics.
\end{thebibliography}
\end{document}